# Bound States in the Continuum in Bilayer Photonic Crystal with TE-TM Cross-Coupling

Hong-Fei Wang,[1] Samit Kumar Gupta,[1] Xue-Yi Zhu,[1] Ming-Hui Lu,[2,*] Xiao-Ping Liu,[2] and Yan-Feng Chen[2]

[1]College of Engineering and Applied Sciences, Nanjing University, Nanjing 210093, China

[2]National Laboratory of Solid State Microstructures, School of Physics and Department of Materials Science and Engineering, Nanjing University, Nanjing 210093, China

Bound states in the continuum (BICs) in photonic crystals represent the unique solutions of wave equations possessing an infinite quality-factor. We design a type of bilayer photonic crystal and study the influence of symmetry and coupling between TE and TM polarizations on BICs. The BIC modes possess $C_{3v}$ symmetry in the x-y plane while the mirror-flip symmetry in the z-direction is broken, and they provide selective coupling into different layers by varying frequency. The enhanced TE-TM coupling due to broken mirror-flip symmetry in the z-direction gives rise to high-Q factor BIC states with unique spatial characteristics. We show the emergence of such BIC states even in the presence of coupling between the TE- and TM-like modes, which is different from the existing single polarization BIC models. We propose to study BICs in multilayer systems, and the results may be helpful in designing photonic settings to observe and manipulate BICs with various symmetries and polarizations for practical applications.

Wave localization draws fundamental as well as technological interests in various areas of physics including optics and photonics, but it poses challenges on the integrated optical circuit (IOC) platforms. Generally, by utilizing structures such as plasmonic microcavities [1] and photonic cavities [2-4], electromagnetic waves can be trapped and the external radiation is suppressed. Until recently, bound states in the continuum (BICs) have attracted substantial interests due to their intriguing nature of localized coexistence with the continuum of leaky radiating modes, which provide an effective mechanism to achieve perfect confinement of light [5-10]. In 1929, von Neumann and Wigner [11] first proposed that there exist unusual solutions for the Schrödinger equation in the quantum system, which correspond to the bound states above the continuum threshold. Subsequently, it was realized that BICs are essentially a kind of wave motion phenomenon. Many different types of BICs have been reported in various physical systems including quantum [11], acoustic waves [12], and electromagnetic waves [13], and so on. More recently, BICs in PhC slabs have been proved to be an ideal platform for studying interesting phenomena due to their inherent ability to tailor the system configurations and material properties. It has been found that waves can be strictly bounded on the slab with a significantly high quality-factor (Q-factor) [5,6], at the frequencies of continuum of unbounded modes.

In addition, we have witnessed growing interest in bilayer graphene [14-18] due to many potentially significant possibilities, including fractional quantum Hall states (FQHs) [19], excitonic condensates [20], and superconductivity in twisted bilayer graphene [21,22], and so on. The bilayer structure provides more design flexibility in terms of rotational and translational operations. Analogous to electronic systems, the bilayer system could be the testbed for intriguing photonic effects as well, such as topological valley transport [23], strong optomechanical coupling [24], and localization [25]. While prior to this study, the investigations on photonic crystals BICs mainly focused on single-layer system configurations such as square [26,27] and triangular [28] lattices, none of these works takes into account the effect of mixed polarizations and complex symmetries reminiscent of multilayer systems.

We propose a rotating bilayer honeycomb photonic crystal slab and study the unique BICs that possess broken mirror-flip symmetry and different polarization components. These BICs exist in photonic crystal slabs, which are dielectric slabs with a periodic modulation of permittivity at the wavelength scale surrounded by air. It is worthwhile to note that this slab structure has been used in many applications, such as surface-emitting lasers [29], biosensing [30,31], and LEDs [32]. Generally, open systems are described by a non-Hermitian effective Hamiltonian [27,34,36], $\mathbf{H} = \mathbf{K} + \mathbf{C_{p1}} + \mathbf{C_{p2}} + \mathbf{C_{rad}}$, where $\mathbf{K}$ corresponds to the contribution of the original uncoupled fundamental waves while the coupling matrices $\mathbf{C_{p1}}$, $\mathbf{C_{p2}}$ and $\mathbf{C_{rad}}$ correspond to the one-dimensional (1D) feedback coupling, 2D coupling via higher order waves, and radiative coupling (see Supplemental Material), possessing multivariate and complex eigenvalues for the modes of the open system. In most cases, these modes correspond to the leaky modes to the surrounding low permittivity medium decaying over time due to the radiation loss. This is evident from theoretical and experimental studies [6,26]. The imaginary parts of the complex eigenfrequencies are used to measure the lifetimes of the corresponding resonator modes. BICs occur when the imaginary parts tend to be zero and the lifetime of the modes approaches infinity.

In this study, as shown in Fig. 1(a), the bilayer PhC slab is surrounded by air. The pump light impinges on it and spreads through the slab. Figure 1(b) shows the cross-section of the 3D system. In this staggered bilayer structure, the symmetry in the x-y plane can be adjusted by rotational and translational operations, while the mirror-flip symmetry in the z-direction is broken. The structure consists of four layers along the z-direction, denoted as Air-$L_1$-$L_2$-Air. The permittivity of air surrounding the slab is 1. We assume that the slab is made of silicon with a permittivity of 11.7 and lattice constant $a$ of 1500 nm. The layers $L_1$ and $L_2$ can be fabricated by engraving periodic holes on a silicon material and wet etching to suspend the structure in air. The structure is considered to be infinite in the x-y plane. The specific lattice structure is shown in Fig. 2. For the lattice constant $a$, the diameters of the two cylindrical holes at lattice sites are $0.2a$ and $0.4a$. We define the effective permittivities of $L_1$ and $L_2$ as $\varepsilon_r^1$ and $\varepsilon_r^2$ respectively with thickness $0.2a$. In the x-y plane, each layer of the slab is a honeycomb PhC with $C_{3v}$ symmetry. Furthermore, between $L_1$ and $L_2$, there is a rotational symmetry such that $D(C_6)L_1 = L_2$, where $D(C_6)$ refers to a $\pi/3$ rotation and $D(C_6) = [1/2, -\sqrt{3}/2; \sqrt{3}/2, 1/2]$. The primitive lattice vectors in the x-y plane of the composite structure are $\boldsymbol{a}_1$ and $\boldsymbol{a}_2$. It should be noted that the PhC slab structure still satisfies $C_{3v}$ symmetry in the x-y plane but it does not maintain mirror-flip symmetry along the z-direction. The breaking of the mirror-flip symmetry will enhance the interaction between TM-like and TE-like modes [13] in the z-direction, which makes it possible to achieve hybrid-polarization BICs.

In our composite system, the mode couplings are enhanced due to the multilayered structure. The band structure contains TM ($H_x, H_y, 0$) and TE-like ($E_x, E_y, 0$) crossings. As a result, the traditional coupled-wave theory (CWT) cannot be applied here [8,9,26,33-36]. The method we propose here is the open boundary coupling theory based on the original Maxwell equations. We assume that the fields are weak enough so that the nonlinear response could be reasonably neglected and the relative magnetic permeability is close to unity. The electric field equation is given by:

$$\boldsymbol{\nabla} \times \boldsymbol{\nabla} \times \mathbf{E}(\mathrm{r}) = \varepsilon_r k_0^2 \mathbf{E}(\mathrm{r}) \tag{1}$$

Here, due to the translational symmetry in the x-y plane, the electric field must satisfy Bloch theorem, $\mathbf{E}(\mathbf{r}_\perp, z) = \mathbf{E}_{\mathbf{k},j}(\mathbf{r}_\perp, z)\exp(i\ \mathbf{k} \cdot \mathbf{r}_\perp)$ and $\mathbf{E}_{\mathbf{k},j}(\mathbf{r}_\perp + \mathbf{R}, z) = \mathbf{E}_{\mathbf{k},j}(\mathbf{r}_\perp, z)$. Here $\mathbf{r}_\perp$ is the position vector in the x-y plane, and $j$ represents the corresponding index number of the solution. $\mathbf{R}$ and $\mathbf{k}$ refer to the in-plane direct lattice vector and Bloch wave vector, respectively. For $L_1$ and $L_2$ layers, the lattice vectors satisfy the condition $\mathbf{R}_{L_1} = \mathbf{R}_{L_2} = \mathbf{R}$. As illustrated in Fig. 1(b), the permittivity of the four layers $\varepsilon_r^0$-$\varepsilon_r^1$-$\varepsilon_r^2$-$\varepsilon_r^0$, can be expressed as $\varepsilon_r^l(\mathbf{r}_\perp + \mathbf{R}, z) = \varepsilon_r^l(\mathbf{r}_\perp, z)$, where $l = 0, 1, 2$. Therefore, we obtain a valid set of equations that efficiently describe the electromagnetic field evolutions in the layers. The finite element method (FEM) is used to solve the above equation. The complex eigenfrequencies of the above equation are defined by $\omega_{\mathbf{k},n} = \omega_{\mathbf{k},n}^0 - i\gamma_{\mathbf{k},n}/2$, with the Q-factors [37-39] given by $Q \triangleq \omega_{\mathbf{k},n}^0/\gamma_{\mathbf{k},n}$. Henceforth, the frequencies and Q-factors of the different BIC modes are discussed. For off-Γ BICs [26], fine-tuning of the system parameters simply shifts the position of these special points along the band diagram, which in effect may limit the stable device prospects. In our PhC slab system, we focus on the band structure in the vicinity of the high-symmetric Γ point.

First, we discuss the symmetry of the BIC modes. These modes at high-symmetric points (Γ) are related to the spatial structure of the slab itself. In order to classify the symmetry of the entire slab in three dimensions, we consider that the mirror-flip symmetry caused by the different size of the holes is perturbed in the z-direction and the structure satisfies $C_{3v}$ symmetry in the x-y plane. Hence, there are six different modes that can exist in this lattice for TE-like or TM-like mode (see Supplemental Material). Figure 3 shows the dispersion relation and Q-factors of the BIC structure along the MΓ and ΓX directions. Modes 1 and 2 in Fig. 3(a) correspond to $(A_1, A')$ and $(A_2, A')$ TE-like modes, respectively, which in turn correspond to the electromagnetic components $(E_x, E_y, 0)$. These BICs are sensitive to symmetry-breaking perturbations as their Q-factors drop sharply away from the Γ point, which essentially points toward the fact that, they are stable as long as the system retains required symmetries. The normalized electric fields of modes 1 and 2 satisfy the x-y plane structural symmetry while being mirror-flip symmetric in the z-direction. However, due to the difference between the upper and lower cylindrical holes, the symmetry of the field distribution is disturbed along the z-direction as shown in Fig. 4. For the TM-like mode with electromagnetic components $(H_x, H_y, 0)$, a similar phenomenon occurs for modes 5 and 6. We describe all four of these modes as the even-like modes since the corresponding overall fields along the z-direction are symmetrical. The field distribution is mainly concentrated at the junction of $L_1$ and $L_2$. As shown in Fig. 3, all four of the above modes are isolated with high Q-factors. This localization behavior stems from the mismatch between the mode profiles residing inside the slab and the external propagating modes. These bound states corresponding to the high-symmetric wave vectors can be used to produce an asymmetrical cavity around an interface.

On the other hand, the electromagnetic fields of the odd modes are mainly concentrated in the central portion of each layer, which can also be considered a dipole in the z-direction. These dipole fields originate from the boundary constraints of the slab in the z-direction. It is worth mentioning here that compared to the even-like modes, the Q-factors for odd-like modes are slightly smaller due to the reduction of energy proportion in the slabs. Notably, modes 7 and 8 possess high Q-factors at the $\Gamma$ point as is evident from Fig. 3(c). Figure 5 shows that mode field distributions in the z-direction resemble the odd symmetry. However, the mismatch between the layers $L_1$ and $L_2$ destroys the perfect mirror symmetry with different field distribution in each of the layers. These two states are expressed as odd-like modes corresponding to $(A_1, B')$ and $(A_2, B')$, respectively. Compared with the earlier reported works on BICs [9,26-28] which primarily focused on monolayer lattices, the modes in our system under consideration have unique symmetry in the x-y plane or in the z-direction, which in effect provides another essential degree of freedom to design the field distribution of the cavity mode. These BICs can be selectively excited at different layer locations by choosing the even or odd-like modes at different frequencies.

Modes 3 and 4 are doubly degenerate while the above-mentioned modes are singly degenerate. Here, these two modes are expressed as $(E, A')$. Since electromagnetic waves admit an E representation in free space, at the $\Gamma$ point, they can only couple to $(E, A')$ modes and have a low Q-factor due to the radiative nature of the layers. It means that the doubly degenerate modes are not BICs by virtue of symmetry-protection since they must be resonance-trapped BICs. Hence, all of the previous modes except modes 3 and 4 are uncoupled by symmetry and possess extremely high Q-factors [40].

From the above symmetry considerations, we would clearly expect that the infinite Q-factor does not rely on the frequency range and single polarization. Moreover, the asymmetric structure raises an interesting possibility for TE-TM cross-coupling. The broken mirror-flip symmetry in the z-direction enhances the coupling between TE and TM modes (see Supplemental Material). Thus, for our bilayer PhC slab, we do not neglect the coupling between TE and TM polarizations. Inevitably, this has led to some unique set of mixed modes that are characteristically present in the TE-TM cross-coupling model. At a slightly higher frequency, the bands of modes 9 and 10 contain the mixed polarizations but still possess the high Q-factor BIC states. The TE-TM cross-coupling occurs in the electromagnetic components $E_x, E_y, E_z \neq 0$ stemming from the cross-coupling between their mode profiles inside the slab and the external propagating modes. This is distinct from the earlier reported BICs [28,36], which essentially points toward the unique fact that, in our system, BICs can be achieved even in the case of TE-TM cross-coupling. The new type of cavity mode with infinite Q-factors and a hybrid TE-TM mode has been proposed in a surface-emitting laser. By filtering the emitted light, TE or TM modes can be selected at will for use in the same system. The in-plane fields of the layers $L_1$ and $L_2$ show that the modes 9 and 10 are in fact higher order modes as is evident from the Fig. 6(b), while in contrast, in the z-direction, they are no longer pure TE-like modes. This is reflected in the angle between the electric field vector and the interface of $L_1$ and $L_2$. The angle at the center of the unit cell even reaches 90°. The energy of these modes is concentrated at the interface, which provides a platform to study local field enhancement in the case of mixed polarizations.

In summary, we have proposed a bilayer honeycomb PhC slab possessing $C_{3v}$ symmetry in the x-y plane and the broken mirror-flip symmetry in the z-direction analogous to electronic systems. We discuss the even-like and odd-like mode field distributions of our bilayer PhC slab. Owing to spatial structure of the layers in this bilayer configuration, this system produces unique spatial BIC modes. More importantly, the broken mirror-flip symmetry in the z-direction enhances the TE-TM coupling resulting in the high Q-factor BIC states. It is observed that in addition to the coupling between single polarization modes and the symmetry of the structure, the coupling between different polarizations also plays a significant role in the formation and emergence of the BICs. Such novel attributes of BICs in the bilayer PhC system could shed further light on BICs in photonic settings in general and in multilayered configurations in particular to develop and broaden our understanding of such localization phenomena and suggest potential applications in manipulating the BIC states in structured photonic platforms for loss-immune communication, lasing, sensing, and so on.

*Acknowledgments*. This work is supported by the National Key R&D Program of China (Grant No. 2017YFA0303700) and the National Nature Science Foundation of China (Grant No. 51721001). We also acknowledge the support of the Academic Program Development of Jiangsu Higher Education (PAPD).

*luminghui@nju.edu.cn

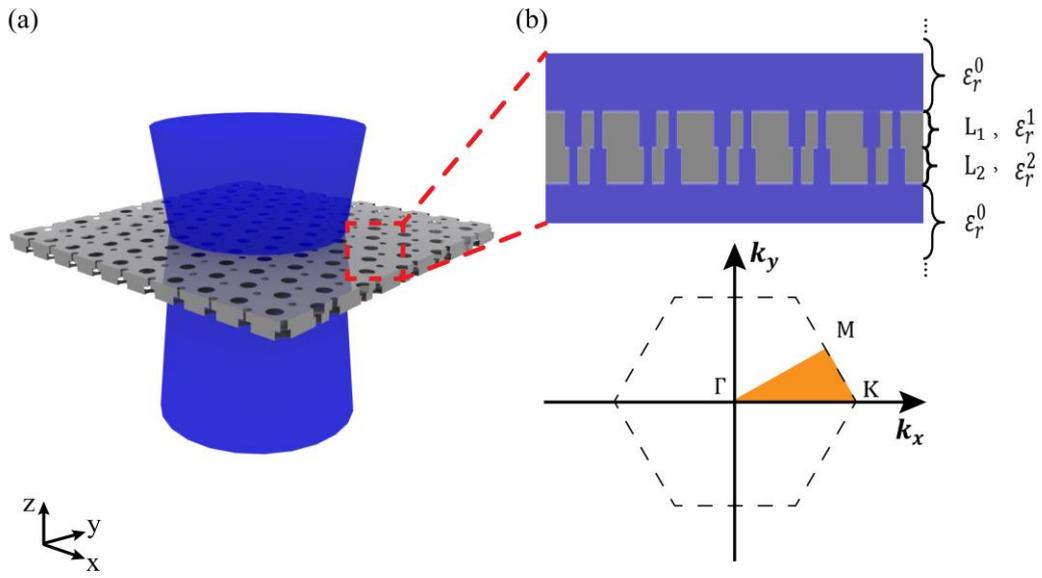

FIG. 1. (a) Schematic of 3D rotating bilayer honeycomb PhC slab fabrication. Blue area represents the pump light excitation and propagation path. (b) Cross-section schematic of PhC slabs, the system is divided into four layers, corresponding to the upper and lower air layers and $L_1$, $L_2$ layers in the middle.

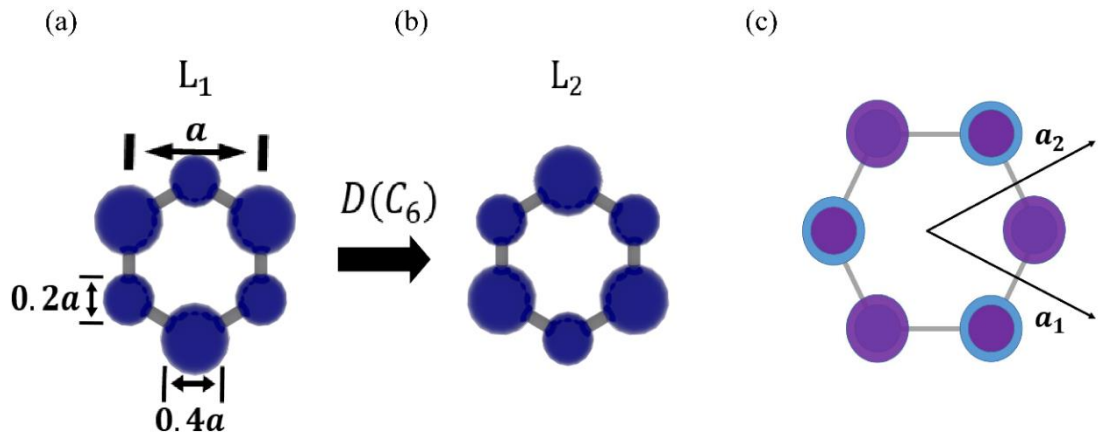

FIG. 2. (a) Schematic of $L_1$ layer unit cell. Lattice constant and air cylindrical holes diameter on site correspond to $a, 0.2a, 0.4a$. (b) $L_2$ layer lattice can be obtained by a $D(C_6)$ rotation of the $L_1$ layer. (c) Structure of two layers together, i.e., the sky-blue and purple layers. $\boldsymbol{a_1}$ and $\boldsymbol{a_2}$ are lattice vectors.

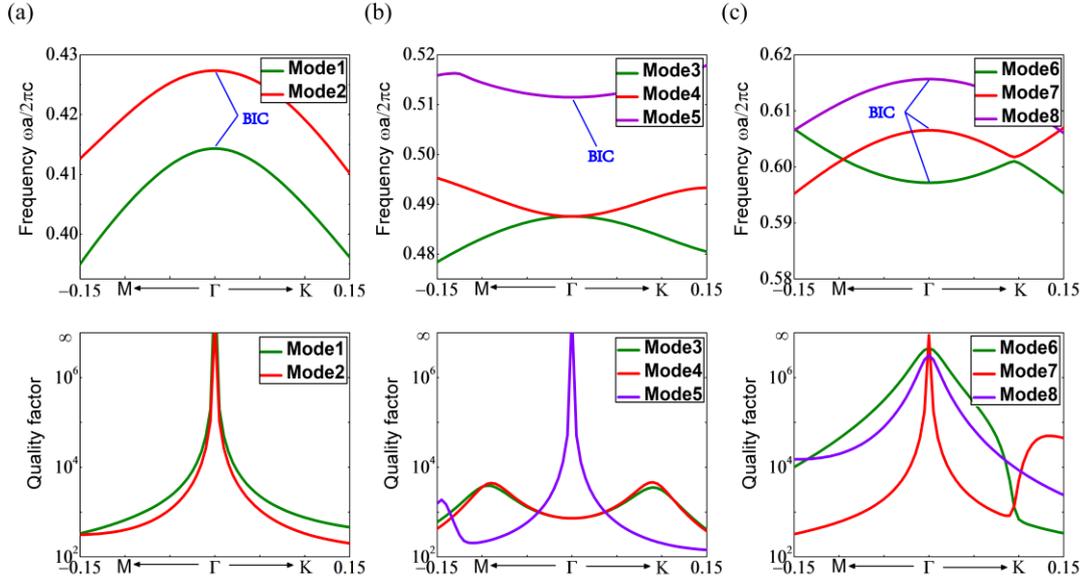

FIG. 3. Band structure and Q-factor of different modes near Γ point along MΓ and ΓK. (a) Modes 1 and 2, with both band structure and Q-factor exhibiting high symmetry. (b) Modes 3, 4 and 5; Only Mode 5 has a considerably high Q-factor around Γ point. (c) Modes 6, 7 and 8.

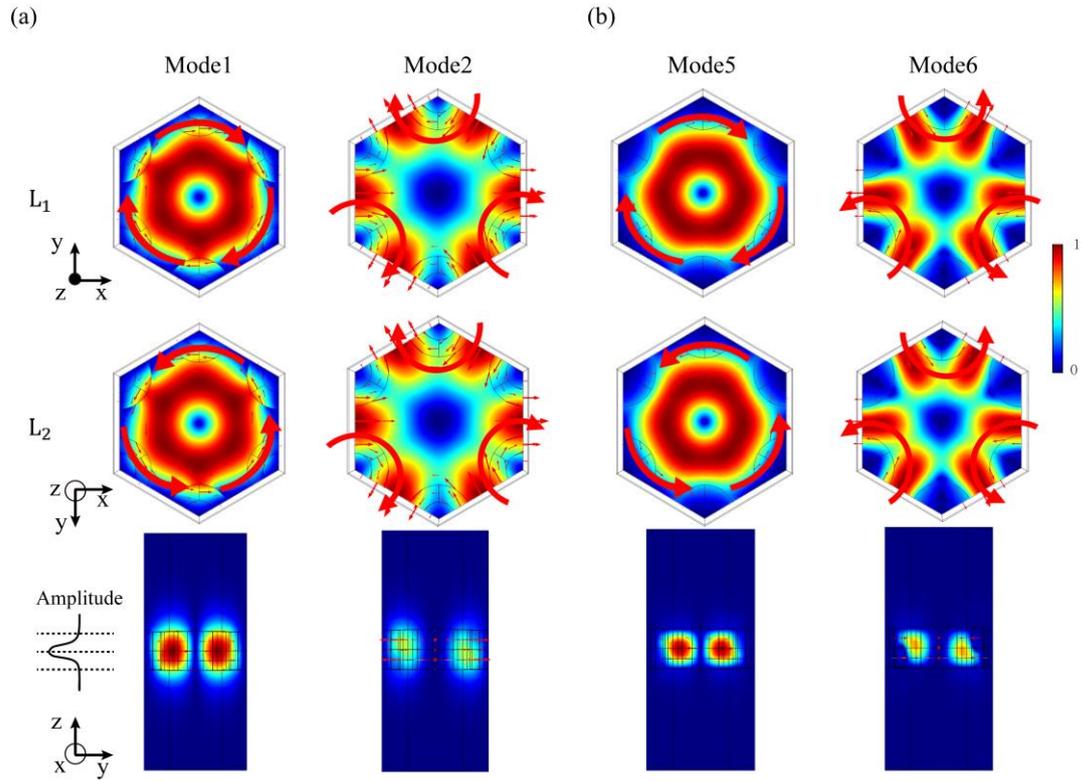

FIG. 4. (a) Normalized BIC electric fields (in color) and E-field vector distribution (red arrows) of modes 1 and 2 in $L_1$, $L_2$ and y-z section. All belong to the TE-like BIC modes and the mirror symmetry is broken in the z-direction. (b) Normalized BIC magnetic fields (in color) and H-field vector (red arrows) of modes 5 and 6 in $L_1$ and $L_2$. Symmetry of the y-z section magnetic field is also broken beyond a distance in the z-direction. Even symmetry of the field distribution is shown in terms of amplitude in the y-z plane plots.

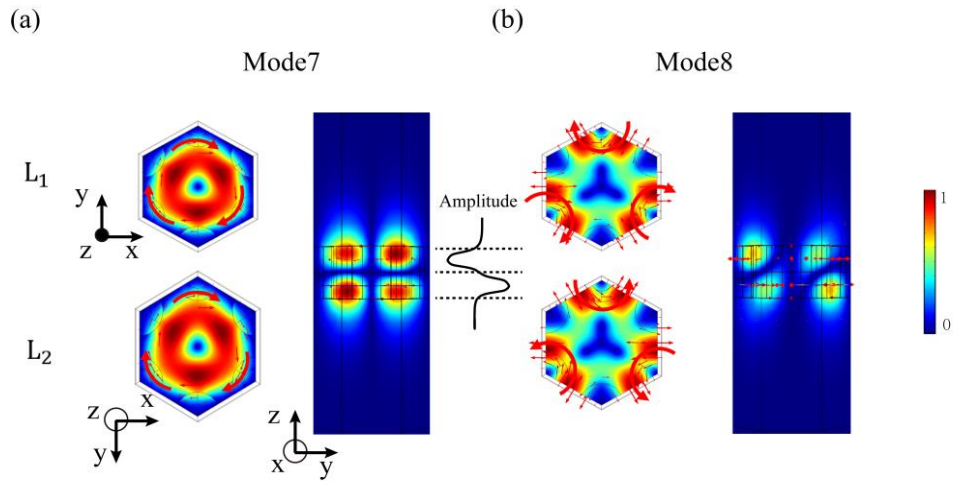

FIG. 5. Normalized BIC electric fields (in color) and electric field vector distribution (red arrows) of different modes. (a) Field distribution of mode 7 in $L_1$, $L_2$. In the y-z section, the electric field is concentrated in the middle of each layer and the mirror symmetry is broken (odd symmetry as shown in terms of amplitude). (b) Field distribution of mode 8 in $L_1$, $L_2$. The y-z section image shows behavior similar to that of mode 7.

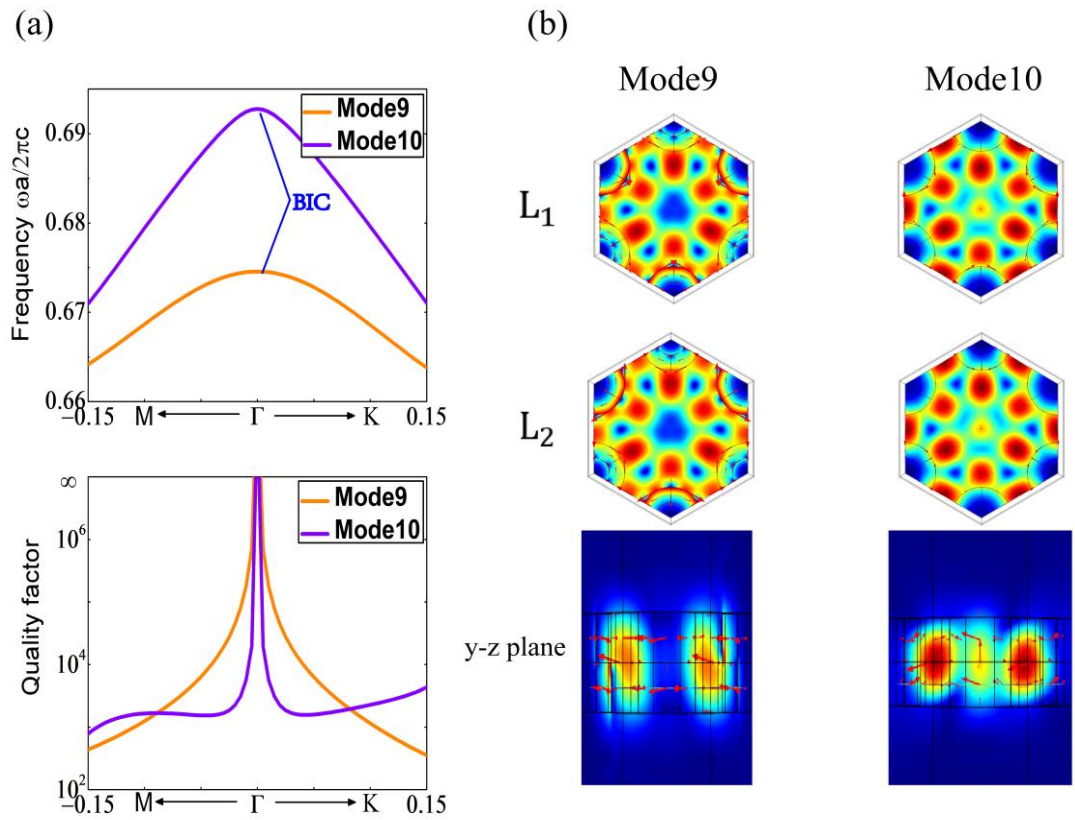

FIG. 6. (a) Higher-frequency band structures and the Q-factors of modes 9 and 10. (b) Normalized BIC electric fields (in color) and E-field vector distribution (red arrows) of modes 9 and 10 in $L_1$, $L_2$.